\documentclass[11pt,a4paper]{article}
\usepackage[latin1]{inputenc}
\usepackage{amsmath}
\usepackage{amsfonts}
\usepackage{amssymb}

\usepackage{geometry}
\usepackage[tt=false]{libertine}
\usepackage[T1]{fontenc}

\usepackage{todonotes}
\usepackage{subcaption}

\title{Reproducibility Report for the Paper:\\\emph{Modeling of Request Cloning in Cloud Server\\Systems using Processor Sharing}}
\author{Alessandro Pellegrini\\pellegrini@diag.uniroma1.it}

\usepackage{tikz,pgfplots,pgfplotstable}
\usepackage{amsmath}
\usepackage{graphicx}
\usepackage{float}

\usepackage{xcolor}
\usepackage{mdframed}

\pgfplotsset{compat=newest}

\usepgfplotslibrary{statistics}
\usetikzlibrary{shapes,arrows, arrows.meta, shapes, positioning}
\usetikzlibrary{pgfplots.statistics}

\pgfplotsset{
	box plot/.style={
		/pgfplots/.cd,
		black,
		only marks,
		mark=-,
		mark size=1em,
		/pgfplots/error bars/.cd,
		y dir=plus,
		y explicit,
	},
	box plot box/.style={
		/pgfplots/error bars/draw error bar/.code 2 args={%
			\draw  ##1 -- ++(1em,0pt) |- ##2 -- ++(-1em,0pt) |- ##1 -- cycle;
		},
		/pgfplots/table/.cd,
		y index=2,
		y error expr={\thisrowno{3}-\thisrowno{2}},
		/pgfplots/box plot
	},
	box plot top whisker/.style={
		/pgfplots/error bars/draw error bar/.code 2 args={%
			\pgfkeysgetvalue{/pgfplots/error bars/error mark}%
			{\pgfplotserrorbarsmark}%
			\pgfkeysgetvalue{/pgfplots/error bars/error mark options}%
			{\pgfplotserrorbarsmarkopts}%
			\path ##1 -- ##2;
		},
		/pgfplots/table/.cd,
		y index=4,
		y error expr={\thisrowno{2}-\thisrowno{4}},
		/pgfplots/box plot
	},
	box plot bottom whisker/.style={
		/pgfplots/error bars/draw error bar/.code 2 args={%
			\pgfkeysgetvalue{/pgfplots/error bars/error mark}%
			{\pgfplotserrorbarsmark}%
			\pgfkeysgetvalue{/pgfplots/error bars/error mark options}%
			{\pgfplotserrorbarsmarkopts}%
			\path ##1 -- ##2;
		},
		/pgfplots/table/.cd,
		y index=5,
		y error expr={\thisrowno{3}-\thisrowno{5}},
		/pgfplots/box plot
	},
	box plot median/.style={
		/pgfplots/box plot
	},
	label style = {font = \small},
	legend style = {font = \footnotesize},
	every axis plot/.append style = {font = \scriptsize}
}

\usepackage{hyperref}

\begin{document}
	
\maketitle

\abstract{
The authors have uploaded their artifact on Zenodo, which ensures a long-term retention of the artifact. The code is suitably documented, and some examples are given. A minimalistic overall description of the engine is provided. The artifact allows to setup the environment quite quickly, and the dependencies are well documented. The process to regenerate data for the figures in the paper completes, and all results are reproducible.

This paper can thus receive the \textit{Artifacts Available} badge and the \textit{Artifacts Evaluated---Functional}. Given the high quality of the artifact, also the \textit{Artifacts Evaluated---Reusable} badge can be assigned.
}

\section{Introduction}

The paper \emph{Modeling of Request Cloning in Cloud Server Systems using Processor Sharing} by Tommi Nylander, Johan Ruuskanen, Karl-Erik Årzén, and Martina Maggio~\cite{paper} presents theoretical results related to the performance of client\slash server applications (in the context of cloud-based deploys) when request cloning is exploited to reduce the time to serve the requests.

The authors present a G/G/1 model which allows to represent a system of servers using cloning to serve the requests as if it were a single server, basing their approach on the idea that the service must be synchronized.

The authors validate their theoretical results by means of a discrete event simulation model, written in python, which is the focus of this report. The authors have also implemented the sequential runtime core for the model, thus making the artifact self-contained, which could be a benefit for some end users, provided they don't have strong performance requirements.

\section{Replication of Computational Results}

In this section we provide Figures taken from the original paper. The authors have given permission to use their original results in this report.

\subsection{Software download and installation}

The authors have provided a link to a permanent repository on Zenodo (\url{http://doi.org/10.5281/zenodo.3611398}), making it permanently available (\textbf{Artifacts Available badge}). This reviewer encourages the authors to place this link also in the main paper, to increase the dissemination of their results.

The amount of dependencies required to run the artifacts is extremely reduced. Simulations can be run by relying solely on python. There is a subsequent post-processing phase, required to analyze the results of the simulations to prepare the datasets to replot the figures which requires matlab. This last step, anyhow, is only related to the reproducibility phase, while the simulation model is self-contained in a reduced number of scripts.

The authors have provided a docker environment to run the simulations, which installs all the required python dependencies, which has been gratefully appreciated.

\subsection{Quality of the artifact}

The artifact is relevant to the associated paper, as it allows to re-generate all the data and all the figures, it is well documented (from the point of view of reproducibility), and it is easy to use thanks to the provided supporting environment. No component to re-run the simulations and to post-process the data is lacking (\textbf{Artifacts Evaluated---Functional}). 

The authors have also provided the results of their simulation runs in the artifact. They propose to rely on this data to regenerate the plots, in case no sufficient computational power is available. I can confirm that by relying on this data, the post-processing and plotting modules do work, and the figures are 1:1 with respect to those in the paper. In the artifact, the authors also provide the post-processed data, thus making it possible to only replot the figures. Matlab scripts are written so that if these post-processed files are found, they are not overwritten. For this reproducibility assessment, this reviewer has rerun all simulations from scratch, and has cleared out all post-processed results before launching the post-processing activity.

The authors have provided a minimalistic documentation to the artifact, consisting in an html file with a high-level description of the simulation model. The authors have also provided a description of the timeline of a cloned request in their software. This documentation does not really allow to repurpose the artifact. Nevertheless, there are some examples which allow to get an insight on how to repurpose the artifact. The model\slash kernel documentation is good, making it easier to understand. Overall, this reviewer thinks that the quality of the artifact is good (\textbf{Artifacts Evaluated---Reusable}).

\subsection{Replicating the experiments}

Replication of the experiments has been carried out on multiple environments. Python simulations have been run on a cluster of 4 parallel high-end machines, all running Debian 9, in the provided docker container. Post processing and plotting have been run on an Arch Linux machine, using matlab R2019b rev~3.

The artifact is easy to run, and all data required to support the reproducibility is provided (\textbf{Artifacts Evaluated---Functional}). It took a more than one week to run all the simulations using $\sim 150$ cores. Initially, two of the post-processing matlab scripts were failing. This was preventing to generate the data which are later required by the tikz scripts in the main latex file to generate the plots.

The authors have been informed via email on Feb 02, 2020 of this issue, receiving all the data from simulations which were re-generated by this reviewer. The authors have replied on Feb 04, 2020, by acknowledging the issues and uploading a revised version of the artifact which has fixed all the aforementioned issues.

In this section, I provide a short description of my reproducibility results, and how they were obtained. There is a total of 11 Figures and 2 Tables in the original paper.

\textbf{Tab. 1} in the paper shows probabilities from a distribution, which does not require to be replicated. Similarly, \textbf{Tab. 2} presents data coming from a theoretical analysis which does not require to be replicated.

\textbf{Fig. 1} in the paper is a drawing and needs not to be reproduced.
\textbf{Fig. 2} in the paper is a graphical representation of multiple equations, thus needs not to be reproduced.

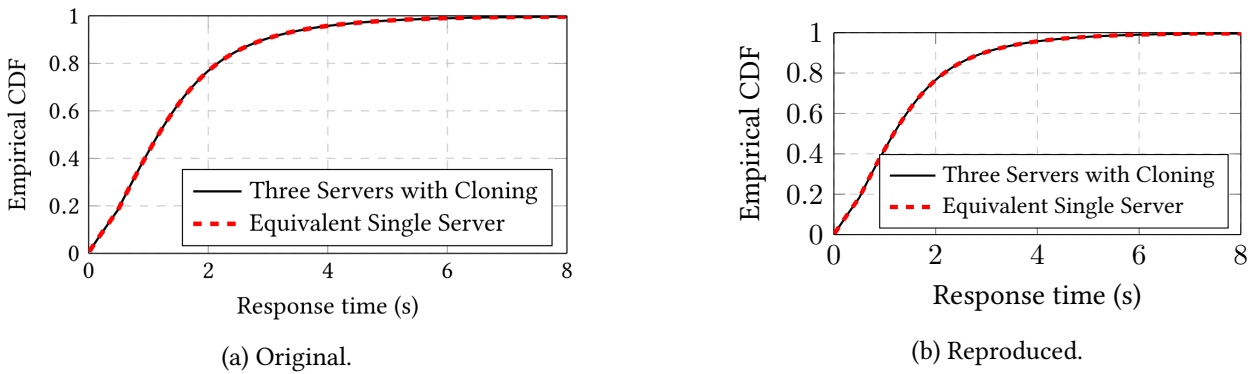
\begin{figure}[h]
	\centering
	\begin{subfigure}{0.45\textwidth}
		\centering
		\includegraphics[width=\textwidth]{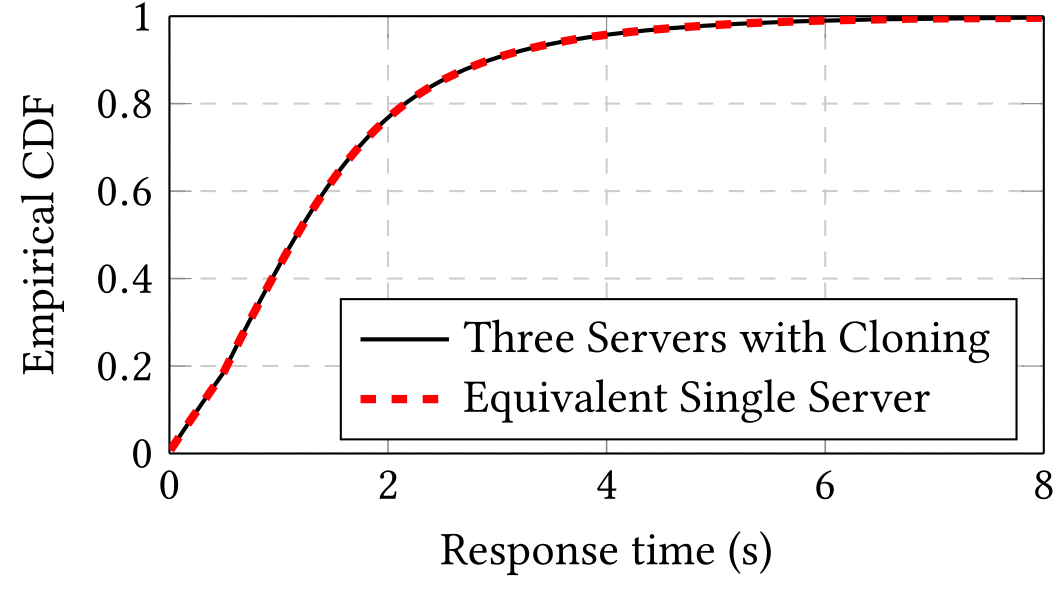}
		\caption{Original.}
	\end{subfigure}
	\hfill
	\begin{subfigure}{0.45\textwidth}
		\centering
		\begin{tikzpicture}[]

\begin{axis}[%
width=0.70\columnwidth,
height=0.35\columnwidth,
scale only axis,
separate axis lines,
xmin=0, xmax=8,
ymin=0, ymax=1,
xlabel = {Response time (s)},
compat = 1.4,
ylabel = {Empirical CDF},
grid style = {dashed, black!20},
grid=major,
legend columns=1,
legend style={draw, fill=white},
legend pos = south east,
ylabel near ticks,
xlabel near ticks,
legend cell align={left},
]

\addplot [color=black, thick, solid]
table [col sep=comma] {data/gg1-example/3dist-ps.csv};
\addlegendentry{Three Servers with Cloning};

\addplot [color=red, ultra thick, dashed]
table [col sep=comma] {data/gg1-example/equivalent-ps.csv};
\addlegendentry{Equivalent Single Server};

\end{axis}
\end{tikzpicture}
		\caption{Reproduced.}
	\end{subfigure}
	\caption{Empirical response time CDFs for the Example with Heterogeneous
		Servers. Data retrieved through 20 repeated simulations of $10^6$ requests each.
		The 95\% confidence intervals lie within the lines.}
	\label{fig:gg1-results-ps}
\end{figure}

\noindent
\textbf{Fig. 3} in the paper corresponds to Figure~\ref{fig:gg1-results-ps} in this report. The data for this figure is generated by running the {\tt sim\_gg1\_3dist.py} wrapper script, which spawns a couple of simulation instances. Data is post processed by the {\tt analyze\_gg1\_example.m} matlab script, which completes successfully. The results are visually and substantially comparable.

\textbf{Fig. 4} in the paper corresponds is a visual representation of a theoretical analysis, which thus does not require reproducibility.
\textbf{Fig. 5} in the paper is a drawing, which does not require reproducibility.

\begin{figure}[h]
	\centering
	\begin{subfigure}{0.45\textwidth}
		\centering
		\includegraphics[width=\textwidth]{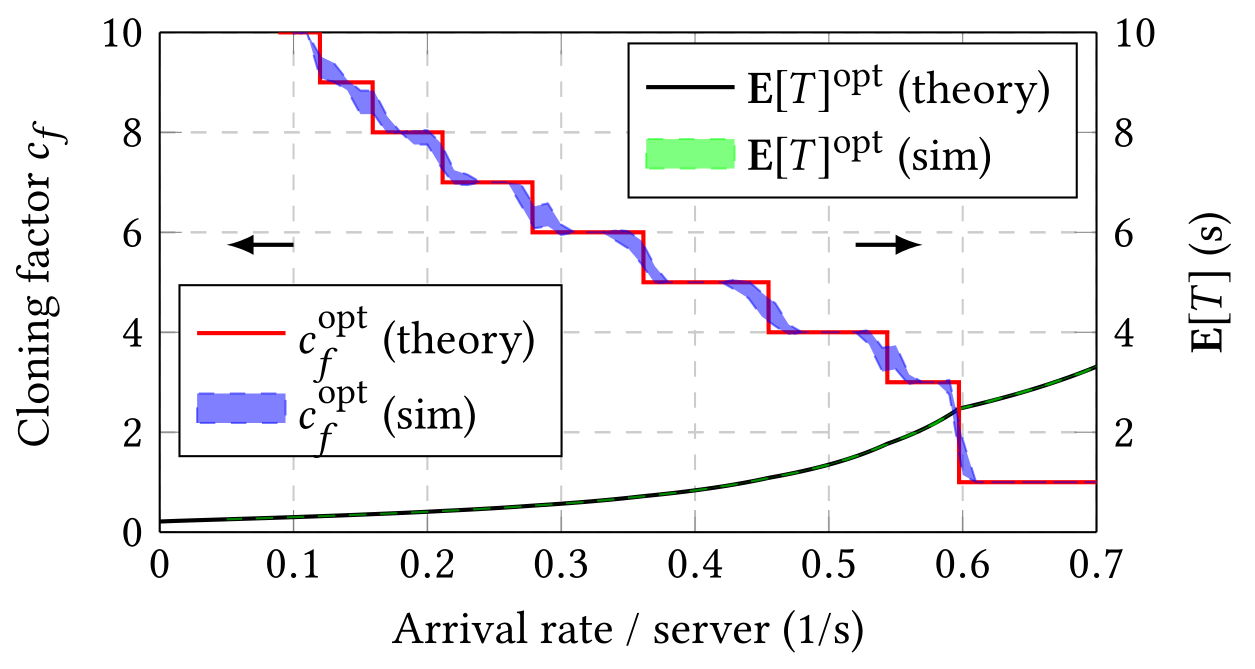}
		\caption{Original.}
	\end{subfigure}
	\hfill
	\begin{subfigure}{0.45\textwidth}
		\centering
		\begin{tikzpicture}[]
\pgfplotsset{set layers}

\begin{axis}[%
width=0.75\columnwidth,
height=0.40\columnwidth,
scale only axis,
separate axis lines,
xmin=0,
xmax=0.7,
ymin=0,
ymax=10,
axis y line*=right,
axis x line=none,
ytick={2, 4, 6, 8, 10},
compat=1.4,
ylabel={$\mathbf{E}[T]$ (s)},
grid style={dashed,black!20},
grid=major,
legend columns=1,
legend style={draw, fill=white, at={(0.98, 0.98)}, anchor=north east},
ylabel near ticks,
xlabel near ticks,
legend cell align={left},
]

\addplot [color=black, thick, solid]
table [col sep=comma] {data/clone-to-all/meanRTs-ps.csv};
\addlegendentry{$\mathbf{E}[T]^{\text{opt}}$ (theory)}


\addplot[area legend,dashed,fill=green,draw=green,opacity=5e-01]
table [col sep=comma]
{data/clone-to-all/optmean-confint.csv};
\addlegendentry{$\mathbf{E}[T]^{\text{opt}}$ (sim)}

\draw[->, >=latex, thick] (axis cs:0.52, 5.75) -- (axis cs:0.57, 5.75);

\end{axis}

\begin{axis}[%
width=0.75\columnwidth,
height=0.40\columnwidth,
scale only axis,
separate axis lines,
axis y line*=left,
xmin=0,
xmax=0.7,
ymin=0,
ymax=10,
xlabel={Arrival rate / server (1/s)},
compat=1.4,
ylabel={Cloning factor $c_f$},
grid style={dashed,black!20},
grid=major,
legend columns=1,
legend style={draw, fill=white, at={(0.02, 0.15)}, anchor=south west},
ylabel near ticks,
xlabel near ticks,
legend cell align={left},
]


\addplot [color=red, thick, solid]
table [col sep=comma] {data/clone-to-all/optclones-ps.csv};
\addlegendentry{$c_f^{\text{opt}}$ (theory)};


\addplot[area legend,dashed,fill=blue,draw=blue,opacity=5e-01]
table [col sep=comma] {data/clone-to-all/optclone-confint.csv};
\addlegendentry{$c_f^{\text{opt}}$ (sim)};



\draw[->, >=latex, thick] (axis cs:0.1, 5.75) -- (axis cs:0.05, 5.75);

\end{axis}

\end{tikzpicture}%
		\caption{Reproduced.}
	\end{subfigure}
	\caption{Clone-to-All: Comparison of theoretical values with 95\% confidence intervals for the simulation results.}
	\label{fig:clone-to-all-evaluation}
\end{figure}
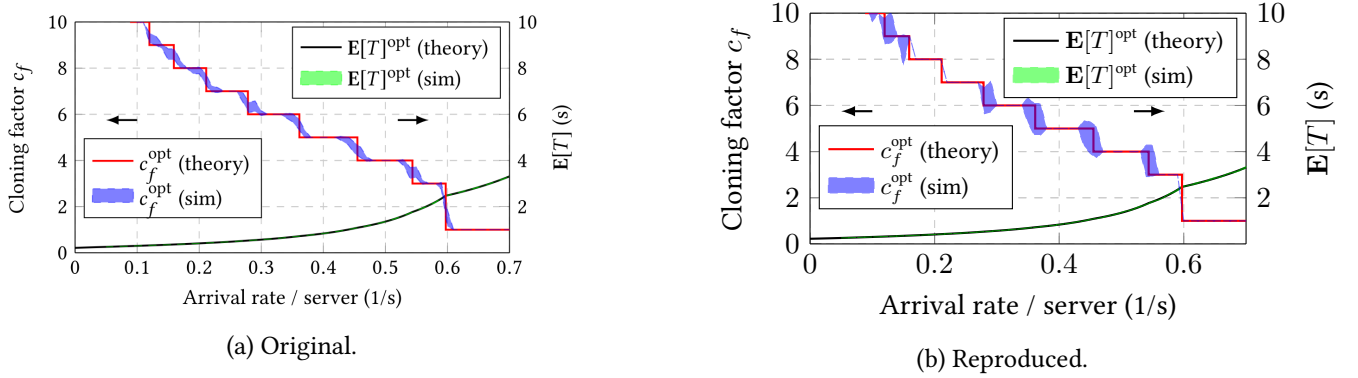

\textbf{Fig. 6} in the paper corresponds to Figure~\ref{fig:clone-to-all-evaluation} in this report.
The data for this figure is generated by running the {\tt sim\_optimal\_clone-ps.py} wrapper script, which spawns 3110 simulation instances. Data is post processed by the {\tt analyze\_opt\_clone.m} matlab script, which completes successfully. The results are slightly different from the original ones in the paper, but the trends are comparable and do not diverge too much. I emphasize that this difference is also related to the reduced number of simulations which have been run for this particular figure, due to a clerical error in the first version of the artifact.

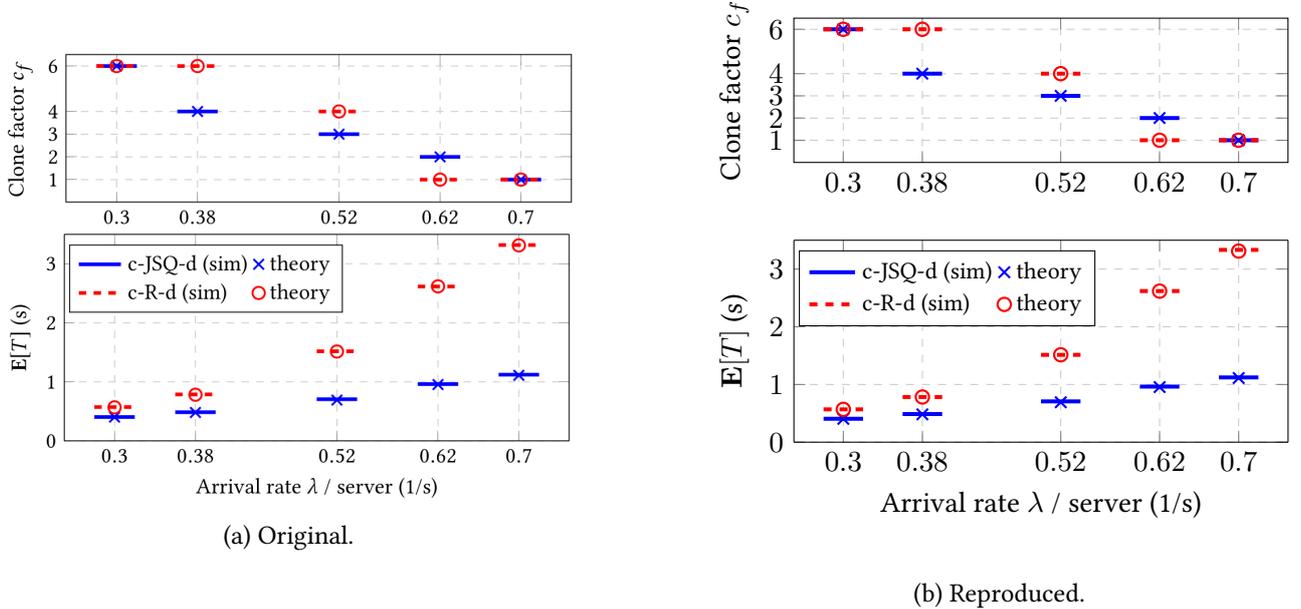
\begin{figure}[h]
	\centering
	\begin{subfigure}{0.45\textwidth}
		\centering
		\includegraphics[width=\textwidth]{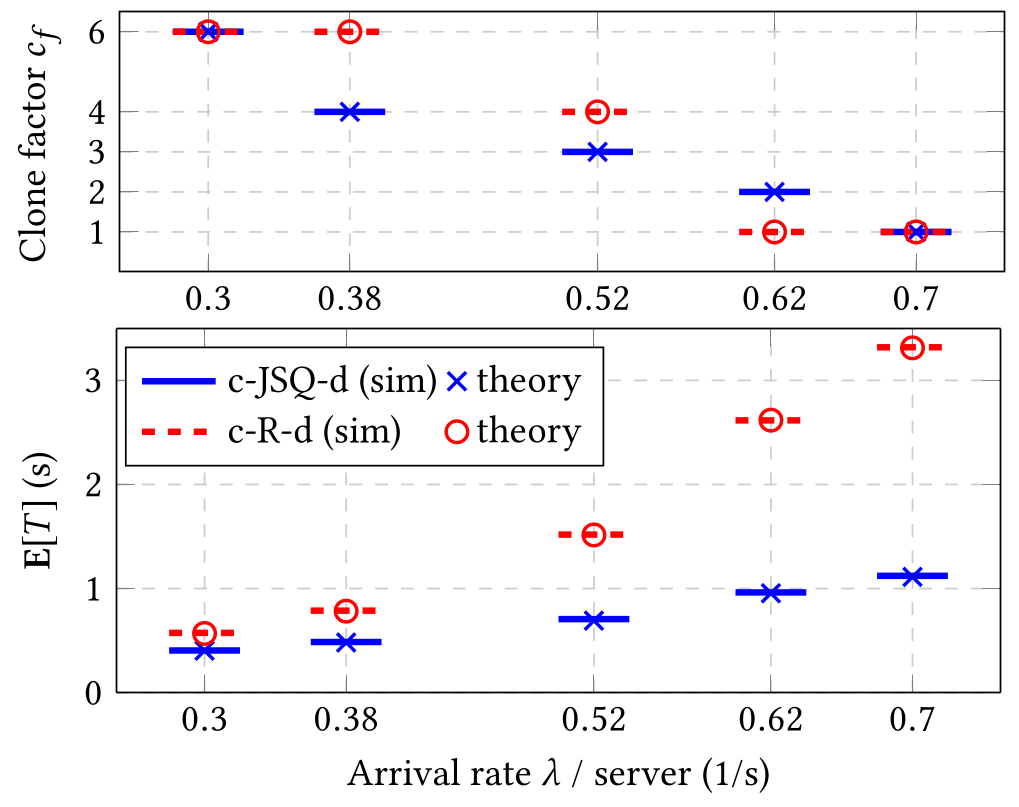}
		\caption{Original.}
	\end{subfigure}
	\hfill
	\begin{subfigure}{0.45\textwidth}
		\centering
		\begin{tikzpicture}

\begin{axis}[%
width=0.85\columnwidth,
height=0.25\columnwidth,
scale only axis,
xmin=0.25,
xmax=0.75,
xtick={0.3, 0.38, 0.52, 0.62, 0.70},
ymin=0,
ymax=6.5,
ytick={1, 2, 3, 4, 6},
ylabel={Clone factor $c_f$},
grid style={dashed,black!20},
grid=major,
legend columns=2,
legend style={draw, fill=white, at={(0.01, 0.05)}, anchor=south west},
ylabel near ticks,
xlabel near ticks,
legend cell align={left},
]

\addplot[solid, blue, line width=1.5pt, restrict expr to domain={\coordindex}{0:1}]
table [col sep=comma] {data/co-design/clusterSQF-PS-clone.csv};

\addplot[blue,mark=x,mark options=solid, only marks,mark size=3pt, thick]
  table[row sep=crcr]{%
x y\\
0.30 6\\
0.38 4\\
0.52 3\\
0.62 2\\
0.70 1\\
};

\addplot[solid, blue, line width=1.5pt, restrict expr to domain={\coordindex}{2:3}, forget plot]
table [col sep=comma] {data/co-design/clusterSQF-PS-clone.csv};

\addplot[solid, blue, line width=1.5pt, restrict expr to domain={\coordindex}{4:5}, forget plot]
table [col sep=comma] {data/co-design/clusterSQF-PS-clone.csv};

\addplot[solid, blue, line width=1.5pt, restrict expr to domain={\coordindex}{6:7}, forget plot]
table [col sep=comma] {data/co-design/clusterSQF-PS-clone.csv};

\addplot[solid, blue, line width=1.5pt, restrict expr to domain={\coordindex}{8:9}, forget plot]
table [col sep=comma] {data/co-design/clusterSQF-PS-clone.csv};

\addplot[dashed, red, line width=1.5pt, restrict expr to domain={\coordindex}{0:1}]
table [col sep=comma] {data/co-design/clusterRandom-PS-clone.csv};

\addplot[red,mark=o, only marks,mark size=2.5pt, thick]
  table[row sep=crcr]{%
x y\\
0.30 6\\
0.38 6\\
0.52 4\\
0.62 1\\
0.70 1\\
};

\addplot[dashed, red, line width=1.5pt, restrict expr to domain={\coordindex}{2:3}, forget plot]
table [col sep=comma] {data/co-design/clusterRandom-PS-clone.csv};

\addplot[dashed, red, line width=1.5pt, restrict expr to domain={\coordindex}{4:5}, forget plot]
table [col sep=comma] {data/co-design/clusterRandom-PS-clone.csv};

\addplot[dashed, red, line width=1.5pt, restrict expr to domain={\coordindex}{6:7}, forget plot]
table [col sep=comma] {data/co-design/clusterRandom-PS-clone.csv};

\addplot[dashed, red, line width=1.5pt, restrict expr to domain={\coordindex}{8:9}, forget plot]
table [col sep=comma] {data/co-design/clusterRandom-PS-clone.csv};

\end{axis}
\end{tikzpicture}
		
		\begin{tikzpicture}

\begin{axis}[%
width=0.85\columnwidth,
height=0.35\columnwidth,
scale only axis,
separate axis lines,
xmin=0.25,
xmax=0.75,
xlabel={Arrival rate $\lambda$ / server (1/s)},
xtick={0.3, 0.38, 0.52, 0.62, 0.70},
ymin=0,
ymax=3.5,
ytick={0,1,2,3},
ylabel={$\mathbf{E}[T]$ (s)},
grid style={dashed,black!20},
grid=major,
legend columns=2,
legend style={draw, fill=white, at={(0.01, 0.95)}, anchor=north west},
ylabel near ticks,
xlabel near ticks,
legend cell align={left},
]

\addplot[solid, blue, line width=1.5pt, restrict expr to domain={\coordindex}{0:1}]
table [col sep=comma] {data/co-design/clusterSQF-PS-RT.csv};
\addlegendentry{c-JSQ-d (sim)}

\addplot[blue,mark=x,mark options=solid, only marks,mark size=3pt, thick]
  table[row sep=crcr]{%
x y\\
0.30 0.404\\
0.38 0.482\\
0.52 0.692\\
0.62 0.955\\
0.70 1.112\\
};
\addlegendentry{theory}

\addplot[solid, blue, line width=1.5pt, restrict expr to domain={\coordindex}{2:3}, forget plot]
table [col sep=comma] {data/co-design/clusterSQF-PS-RT.csv};

\addplot[solid, blue, line width=1.5pt, restrict expr to domain={\coordindex}{4:5}, forget plot]
table [col sep=comma] {data/co-design/clusterSQF-PS-RT.csv};

\addplot[solid, blue, line width=1.5pt, restrict expr to domain={\coordindex}{6:7}, forget plot]
table [col sep=comma] {data/co-design/clusterSQF-PS-RT.csv};

\addplot[solid, blue, line width=1.5pt, restrict expr to domain={\coordindex}{8:9}, forget plot]
table [col sep=comma] {data/co-design/clusterSQF-PS-RT.csv};

\addplot[dashed, red, line width=1.5pt, restrict expr to domain={\coordindex}{0:1}]
table [col sep=comma] {data/co-design/clusterRandom-PS-RT.csv};
\addlegendentry{c-R-d (sim)}

\addplot[red,mark=o, only marks,mark size=2.5pt, thick]
  table[row sep=crcr]{%
x y\\
0.30 0.569\\
0.38 0.783\\
0.52 1.515\\
0.62 2.618\\
0.70 3.312\\
};
\addlegendentry{theory}

\addplot[dashed, red, line width=1.5pt, restrict expr to domain={\coordindex}{2:3}, forget plot]
table [col sep=comma] {data/co-design/clusterRandom-PS-RT.csv};

\addplot[dashed, red, line width=1.5pt, restrict expr to domain={\coordindex}{4:5}, forget plot]
table [col sep=comma] {data/co-design/clusterRandom-PS-RT.csv};

\addplot[dashed, red, line width=1.5pt, restrict expr to domain={\coordindex}{6:7}, forget plot]
table [col sep=comma] {data/co-design/clusterRandom-PS-RT.csv};

\addplot[dashed, red, line width=1.5pt, restrict expr to domain={\coordindex}{8:9}, forget plot]
table [col sep=comma] {data/co-design/clusterRandom-PS-RT.csv};
\end{axis}
\end{tikzpicture}
		\caption{Reproduced.}
	\end{subfigure}
	\caption{Co-designs: Comparison of theoretical values with 95\% confidence intervals for the simulation results. The legend applies to both figures.}
	\label{fig:co-design}
\end{figure}

\textbf{Fig. 7} in the paper corresponds to Figure~\ref{fig:co-design} in this report.
The data for this figure is generated by running the {\tt sim\_codesigns\_icpe.py} wrapper script, which spawns 880 simulation instances. Data is post processed by the {\tt analyze\_codesigns.m} matlab script, which completes successfully. The results are visually and substantially comparable.

\begin{figure}[h]
	\centering
	\begin{subfigure}{0.45\textwidth}
		\centering
		\includegraphics[width=\textwidth]{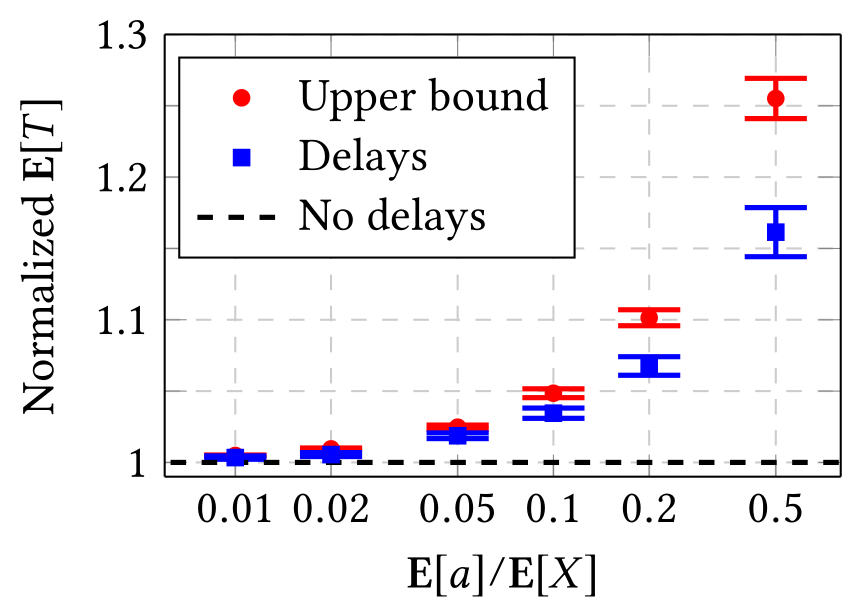}
		\caption{Original.}
	\end{subfigure}
	\hfill
	\begin{subfigure}{0.45\textwidth}
		\centering
		\begin{tikzpicture}

\begin{axis}[%
width=0.82\columnwidth,
height=3.0cm,
scale only axis,
separate axis lines,
xmin=0.006,
xmax=0.8,
xlabel={$\frac{\mathbf{E}[a]}{\mathbf{E}[X]}$},
ylabel={Normalized $\mathbf{E}[T]$},
xmode=log,
xtick = data,
log ticks with fixed point,
ymin=0.99,
ymax=1.3,
ytick={1, 1.05, 1.1, 1.15, 1.20, 1.25, 1.3},
yticklabels = {1,, 1.1,, 1.2,, 1.3},
compat=1.5.1,
grid style={dashed,black!20},
grid=major,
legend columns=1,
legend style={draw, fill=white, at={(0.02, 0.95)}, anchor=north west},
ylabel near ticks,
xlabel near ticks,
legend cell align={left},
]

\addplot+[red, only marks, mark options={red}, mark size=1.5pt,
  error bars/.cd,
    y dir=both,
    y explicit,
    error bar style={line width=1pt},
    error mark options={
      rotate=90,
      mark size=6pt,
      line width=1pt
    }
]
table [x index=0, y index=2, x error index=1, y error index=3] {data/randomized-delays/randomized_arrival_delays_confint_bound.txt};
\addlegendentry{Upper bound}

\addplot+[blue, only marks, mark options={blue}, mark size=1.5pt,
  error bars/.cd,
    y dir=both,
    y explicit,
    error bar style={line width=1pt},
    error mark options={
      rotate=90,
      mark size=6pt,
      line width=1pt
    }
]
table [x index=0, y index=2, x error index=1, y error index=3] {data/randomized-delays/randomized_arrival_delays_confint_resp.txt};
\addlegendentry{Delays}

\addplot[dashed, black, line width=1.0pt]
  table[row sep=crcr]{%
x y\\
0.005 1\\
0.8 1\\
};
\addlegendentry{No delays}

\end{axis}
\end{tikzpicture}
		\caption{Reproduced.}
	\end{subfigure}
	\caption{Arrival delays only.}
	\label{8a}
\end{figure}
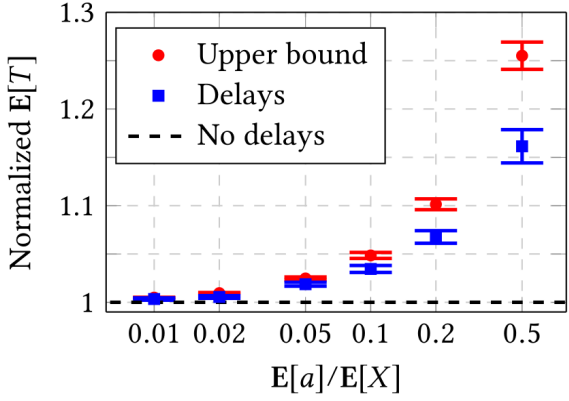
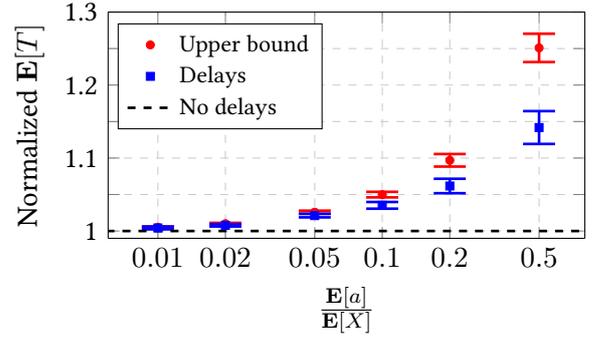

\textbf{Fig. 8(a)} in the paper corresponds to Figure~\ref{8a} in this report.
The data for this figure is generated by running the {\tt sim\_randomized\_arrival\_delays.py} wrapper script, which spawns 1000 simulation instances. Data is post processed by the {\tt analyze\_randomized\_arrival\_delays.m} matlab script, which completes successfully. The results are visually and substantially comparable, except for the point at $x=0.5$, but the deviation can be deemed as negligible.

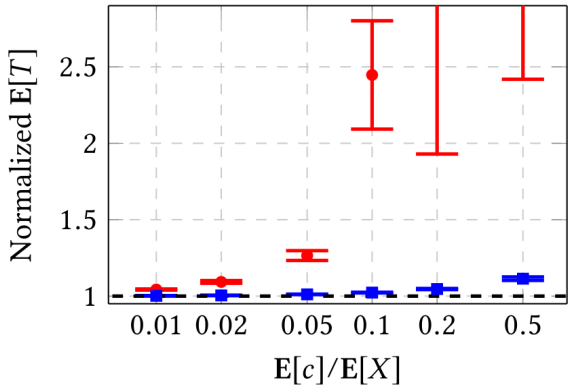
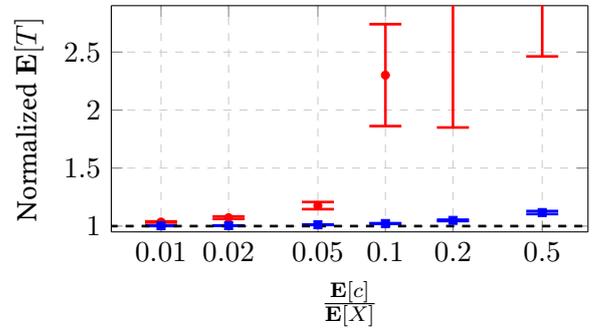
\begin{figure}[h]
	\centering
	\begin{subfigure}{0.45\textwidth}
		\centering
		\includegraphics[width=\textwidth]{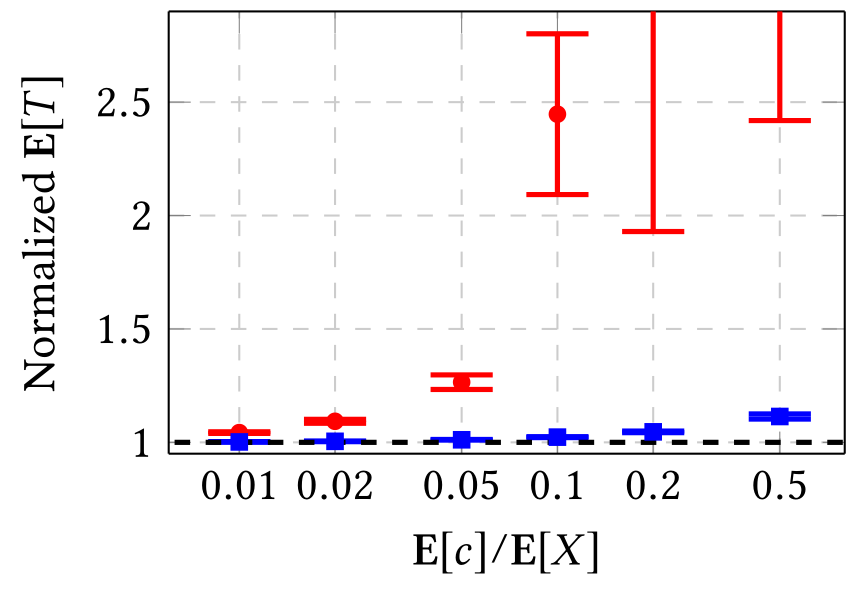}
		\caption{Original.}
	\end{subfigure}
	\hfill
	\begin{subfigure}{0.45\textwidth}
		\centering
		\begin{tikzpicture}

\begin{axis}[%
width=0.82\columnwidth,
height=3.0cm,
scale only axis,
separate axis lines,
xmin=0.006,
xmax=0.8,
xlabel={$\frac{\mathbf{E}[c]}{\mathbf{E}[X]}$},
ylabel={Normalized $\mathbf{E}[T]$},
xmode=log,
xtick = data,
log ticks with fixed point,
ymin=0.95,
ymax=2.90,
compat=1.5.1,
grid style={dashed,black!20},
grid=major,
legend columns=1,
legend style={draw, fill=white, at={(0.02, 0.95)}, anchor=north west},
ylabel near ticks,
xlabel near ticks,
legend cell align={left},
]

\addplot+[red, only marks, mark options={red}, mark size=1.5pt,
  error bars/.cd,
    y dir=both,
    y explicit,
    error bar style={line width=1pt},
    error mark options={
      rotate=90,
      mark size=6pt,
      line width=1pt
    }
]
table [x index=0, y index=2, x error index=1, y error index=3] {data/randomized-delays/randomized_cancellation_delays_confint_bound.txt};

\addplot+[blue, only marks, mark options={blue}, mark size=1.5pt,
  error bars/.cd,
    y dir=both,
    y explicit,
    error bar style={line width=1pt},
    error mark options={
      rotate=90,
      mark size=6pt,
      line width=1pt
    }
]
table [x index=0, y index=2, x error index=1, y error index=3] {data/randomized-delays/randomized_cancellation_delays_confint_resp.txt};

\addplot[dashed, black, line width=1.0pt]
  table[row sep=crcr]{%
x y\\
0.005 1\\
0.8 1\\
};

\end{axis}
\end{tikzpicture}
		\caption{Reproduced.}
	\end{subfigure}
	\caption{Cancellation delays only.}
	\label{8b}
\end{figure}

\textbf{Fig. 8(b)} in the paper corresponds to Figure~\ref{8b} in this report.
The data for this figure is generated by running the {\tt sim\_randomized\_cancellation\_delays.py} wrapper script, which spawns 1000 simulation instances. Data is post processed by the {\tt analyze\_randomized\_cancellation\_delays.m} matlab script, which completes successfully. The results are visually and substantially comparable, except for the point at $x=0.1$, but the deviation can be deemed as negligible.

\begin{figure}[h]
	\centering
	\begin{subfigure}{0.45\textwidth}
		\centering
		\includegraphics[width=\textwidth]{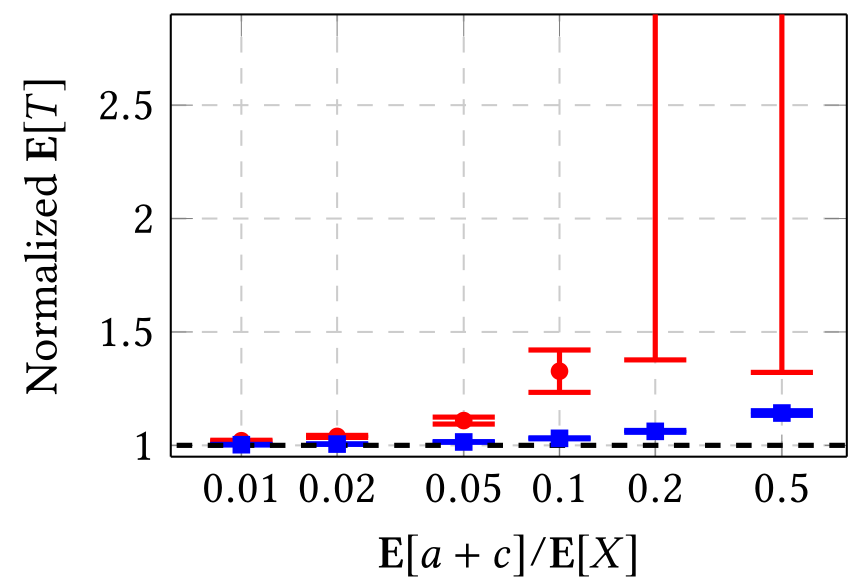}
		\caption{Original.}
	\end{subfigure}
	\hfill
	\begin{subfigure}{0.45\textwidth}
		\centering
		\begin{tikzpicture}

\begin{axis}[%
width=0.82\columnwidth,
height=3.0cm,
scale only axis,
separate axis lines,
xmin=0.006,
xmax=0.8,
xlabel={$\frac{\mathbf{E}[a+c]}{\mathbf{E}[X]}$},
ylabel={Normalized $\mathbf{E}[T]$},
xmode=log,
xtick = data,
log ticks with fixed point,
ymin=0.95,
ymax=2.90,
compat=1.5.1,
grid style={dashed,black!20},
grid=major,
legend columns=1,
legend style={draw, fill=white, at={(0.02, 0.95)}, anchor=north west},
ylabel near ticks,
xlabel near ticks,
legend cell align={left},
]

\addplot+[red, only marks, mark options={red}, mark size=1.5pt,
  error bars/.cd,
    y dir=both,
    y explicit,
    error bar style={line width=1pt},
    error mark options={
      rotate=90,
      mark size=6pt,
      line width=1pt
    }
]
table [x index=0, y index=2, x error index=1, y error index=3] {data/randomized-delays/randomized_combined_delays_confint_bound.txt};

\addplot+[blue, only marks, mark options={blue}, mark size=1.5pt,
  error bars/.cd,
    y dir=both,
    y explicit,
    error bar style={line width=1pt},
    error mark options={
      rotate=90,
      mark size=6pt,
      line width=1pt
    }
]
table [x index=0, y index=2, x error index=1, y error index=3] {data/randomized-delays/randomized_combined_delays_confint_resp.txt};

\addplot[dashed, black, line width=1.0pt]
  table[row sep=crcr]{%
x y\\
0.005 1\\
0.8 1\\
};

\end{axis}
\end{tikzpicture}
		\caption{Reproduced.}
	\end{subfigure}
	\caption{Both delays present.}
	\label{8c}
\end{figure}
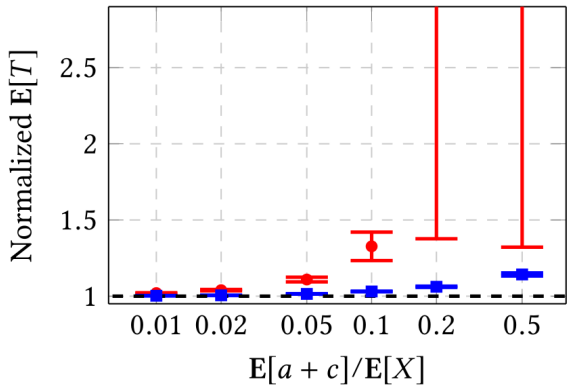
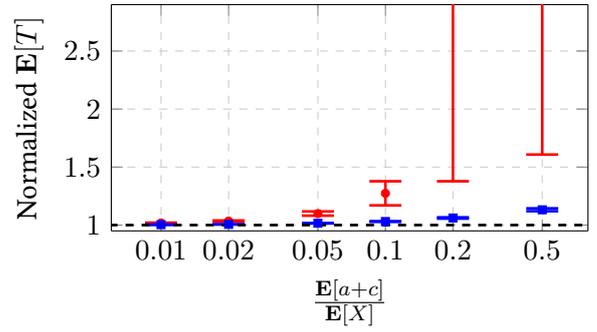

\noindent
\textbf{Fig. 8(c)} in the paper corresponds to Figure~\ref{8c} in this report.
The data for this figure is generated by running the {\tt sim\_randomized\_combined\_delays.py} wrapper script, which spawns 1000 simulation instances. Data is post processed by the {\tt analyze\_randomized\_combined\_delays.m} matlab script, which completes successfully, except for the error bar at $x=0.05$, but the deviation can be deemed as negligible and the trend might look overall better than the original plot in the paper.

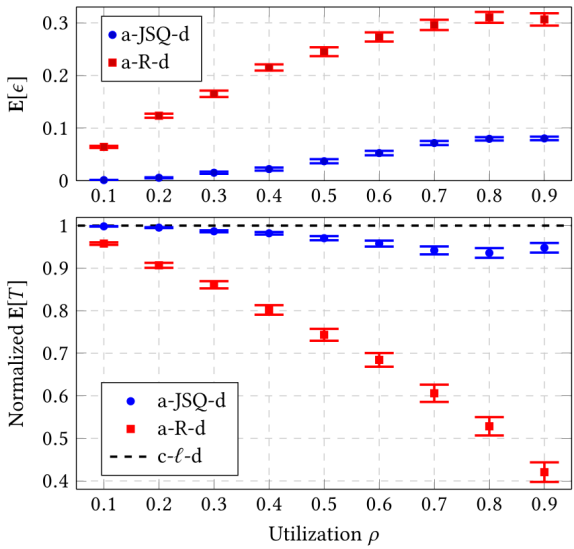
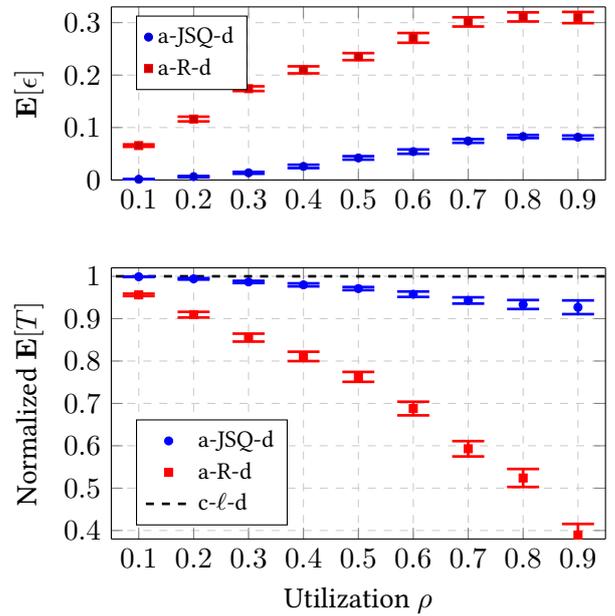
\begin{figure}[h]
	\centering
	\begin{subfigure}{0.45\textwidth}
		\centering
		\includegraphics[width=\textwidth]{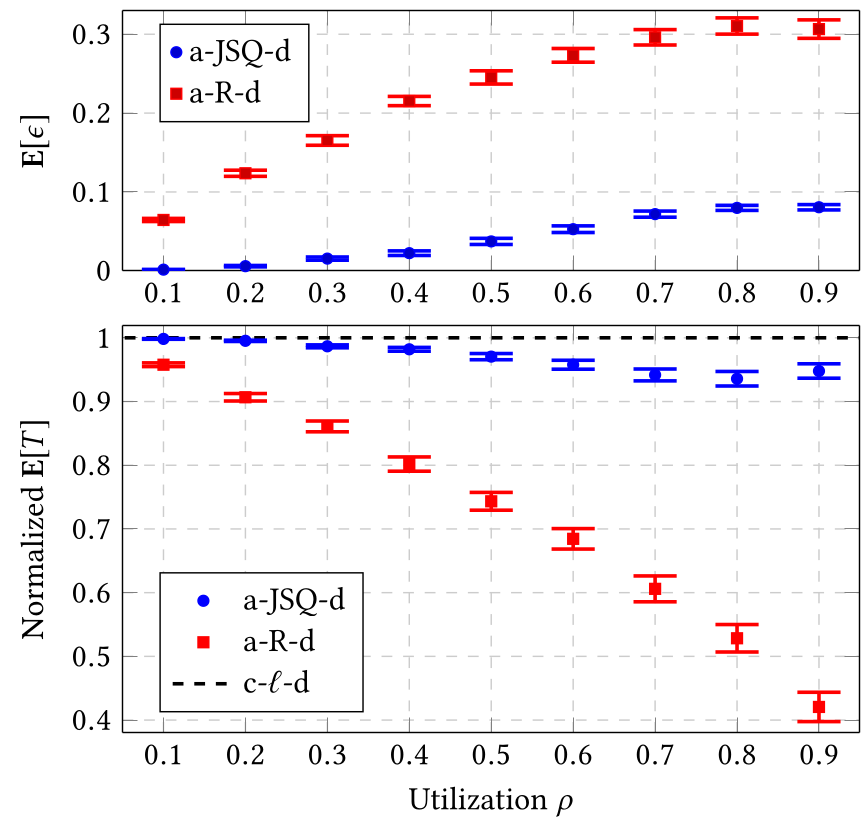}
		\caption{Original.}
	\end{subfigure}
	\hfill
	\begin{subfigure}{0.45\textwidth}
		\centering
		\begin{tikzpicture}

\begin{axis}[%
width=0.85\columnwidth,
height=0.30\columnwidth,
scale only axis,
separate axis lines,
xmin=0.05,
xmax=0.95,
xtick={0.1, 0.2, 0.3, 0.4, 0.5, 0.6, 0.7, 0.8, 0.9},
ymin=0.0,
ymax=0.33,
compat=1.5.1,
ylabel={$\mathbf{E}[\epsilon]$},
grid style={dashed,black!20},
grid=major,
legend columns=1,
legend style={draw, fill=white, at={(0.05, 0.95)}, anchor=north west},
ylabel near ticks,
xlabel near ticks,
legend cell align={left},
]

\addplot+[color=blue, only marks, mark=*, mark size=1.5pt,
  error bars/.cd,
    y dir=both,
    y explicit,
    error bar style={line width=1pt},
    error mark options={
      rotate=90,
      mark size=6pt,
      line width=1pt
    }
]
table [x index=0, y index=2, x error index=1, y error index=3] {data/randomized-sync-vs-nonsync/randomized_sqf_clone_confint.txt};
\addlegendentry{a-JSQ-d}

\addplot+[color=red, only marks, mark=square*, mark size=1.5pt,
  error bars/.cd,
    y dir=both,
    y explicit,
    error bar style={line width=1pt},
    error mark options={
      rotate=90,
      mark size=6pt,
      line width=1pt
    }
]
table [x index=0, y index=2, x error index=1, y error index=3] {data/randomized-sync-vs-nonsync/randomized_random_clone_confint.txt};
\addlegendentry{a-R-d}

\end{axis}
\end{tikzpicture}
		
		\begin{tikzpicture}

\begin{axis}[%
width=0.85\columnwidth,
height=0.47\columnwidth,
scale only axis,
separate axis lines,
xmin=0.05,
xmax=0.95,
xtick={0.1, 0.2, 0.3, 0.4, 0.5, 0.6, 0.7, 0.8, 0.9},
ymin=0.38,
xlabel={Utilization $\rho$},
ymax=1.02,
ytick={0.4, 0.5, 0.6, 0.7, 0.8, 0.9, 1},
compat=1.5.1,
ylabel={Normalized $\mathbf{E}[T]$},
grid style={dashed,black!20},
grid=major,
legend columns=1,
legend style={draw, fill=white, at={(0.05, 0.05)}, anchor=south west},
ylabel near ticks,
reverse legend,
xlabel near ticks,
legend cell align={left},
]

\addplot[dashed, black, line width=1.0pt]
  table[row sep=crcr]{%
x y\\
0.0 1\\
1.0 1\\
};
\addlegendentry{c-$\ell$-d}

\addplot+[red, only marks, mark options={red}, mark size=1.5pt,
  error bars/.cd,
    y dir=both,
    y explicit,
    error bar style={line width=1pt},
    error mark options={
      rotate=90,
      mark size=6pt,
      line width=1pt
    }
]
table [x index=0, y index=2, x error index=1, y error index=3] {data/randomized-sync-vs-nonsync/randomized_random_mean_confint.txt};
\addlegendentry{a-R-d}

\addplot+[blue, only marks, mark options={blue}, mark size=1.5pt,
  error bars/.cd,
    y dir=both,
    y explicit,
    error bar style={line width=1pt},
    error mark options={
      rotate=90,
      mark size=6pt,
      line width=1pt
    }
]
table [x index=0, y index=2, x error index=1, y error index=3] {data/randomized-sync-vs-nonsync/randomized_sqf_mean_confint.txt};
\addlegendentry{a-JSQ-d}

\end{axis}
\end{tikzpicture}
		\caption{Reproduced.}
	\end{subfigure}
		\caption{Simulations comparing a-$\ell$-d to c-$\ell$-d for random and JSQ. The normalization of $\mathbf{E}[T]$ is performed such that each value is divided by the value for the c-$\ell$-d counterpart. The intervals represent 95\% confidence intervals.}
	\label{fig:nonsync-vs-sync}
\end{figure}

\noindent
\textbf{Fig. 9} in the paper corresponds to Figure~\ref{fig:nonsync-vs-sync} in this report.
The data for this figure is generated by running the {\tt sim\_randomized\_sync\_vs\_nonsync\_icpe.py} wrapper script, which spawns 4000 simulation instances. Data is post processed by the {\tt analyze\_randomized\_scenarios.m} matlab script, which completes successfully, except for the points at $x=0.9$, but the deviation can be deemed as negligible and the trend might look overall better than the original plot in the paper.

Overall, the plots which have been generated are consistent with the original publication.


\begin{thebibliography}{1}
	\bibitem{paper}Tommi Nylander, Johan Ruuskanen, Karl-Erik Årzén, and Martina Maggio,
	``Modeling of Request Cloning in Cloud Server Systems using Processor Sharing,"
	In~\emph{Proceedings of the 11$^{th}$ ACM/SPEC International Conference on Performance Engineering}, April 2020.
\end{thebibliography}
\end{document}